\documentclass{elsarticle}
\pdfoutput=1
\usepackage{amsopn}

\newcommand{\IR}{\mathbb{R}}

\newcommand{\CC}{\mathcal{C}}

\newcommand{\BB}{\mathcal{B}}
\newcommand{\IE}{\mathscr{I}}
\newcommand{\MM}{\mathcal{M}}
\newcommand{\IM}{\mathscr{M}}

\newcommand{\ff}{\mathcal{F}}
\newcommand{\IF}{\mathscr{F}}

\newcommand{\IA}{\mathbb{M}}

\newcommand{\HH}{\eta}

\newcommand{\Domain}{\mathscr{D}}
\newcommand{\EQ}[1]{(\ref{eq:#1})}
\newcommand{\SEC}[1]{\ref{sec:#1}}
\newcommand{\FIG}[1]{\ref{fig:#1}}
\newcommand{\feq}{f_{\mathrm{eq}}}
\newcommand{\llbrack}{{\lbrack\hspace{-1.25pt}\lbrack}}
\newcommand{\rrbrack}{{\rbrack\hspace{-1.25pt}\rbrack}}
\newcommand{\mean}[1]{\{\!\!\{ #1 \}\!\!\}}
\newcommand{\jump}[1]{\llbrack #1 \rrbrack}
\newcommand{\bjump}[1]{\big[\hspace{-2.5pt}\big[{#1}\big]\hspace{-2.5pt}\big]}
\newcommand{\Kn}{\mathrm{Kn}}

\newcommand*{\defeq}{\mathrel{\rlap{%
                     \raisebox{0.3ex}{$\m@th\cdot$}}%
                     \raisebox{-0.3ex}{$\m@th\cdot$}}%
                     =}
\newcommand*{\eqdef}{=\mathrel{\rlap{%
                     \raisebox{0.3ex}{$\m@th\cdot$}}%
                     \raisebox{-0.3ex}{$\m@th\cdot$}}%
                     }

\DeclareMathOperator{\argmin}{arg\:min}

\usepackage{amsmath}
\usepackage{amsthm}
\usepackage{mathrsfs}
\usepackage{amssymb}
\usepackage{bm}
\usepackage[margin=2.3cm]{geometry}
\usepackage{color}

\newtheorem*{remark}{Remark}

\usepackage{citesort}

\begin{document}

\begin{frontmatter}

\title{An Entropy Stable Discontinuous Galerkin Finite-Element Moment Method for the Boltzmann Equation} 

\author{M.R.A. Abdelmalik\corref{mycorrespondingauthor}} \author{E.H. van Brummelen}  %
\address{Department of Mechanical Engineering, Eindhoven University of Technology, The Netherlands}
\cortext[mycorrespondingauthor]{Corresponding author}
\ead{m.abdel.malik@tue.nl}

\begin{abstract}
This paper presents a numerical approximation technique for the Boltzmann equation based on a moment-system
approximation in velocity dependence and a discontinuous Galerkin finite-element approximation in position dependence. The closure relation for the moment systems derives from minimization of a suitable $\varphi$-divergence. This divergence-based closure yields a hierarchy of tractable symmetric hyperbolic moment systems that retain the fundamental structural properties of the Boltzmann equation. The resulting combined discontinuous Galerkin moment method corresponds to a Galerkin approximation of the
Boltzmann equation in renormalized form. We present a new class of numerical flux functions, based on the underlying renormalized Boltzmann equation, that ensure entropy dissipation of the approximation scheme. Numerical results are presented for a one-dimensional test case.
\end{abstract}


\end{frontmatter}


\section{Introduction}
\label{sec:intro}
The Boltzmann equation provides a description of the molecular dynamics of fluid flows based on their one-particle phase-space distribution. However,
the Boltzmann equation also encapsulates all conventional macroscopic flow models in the sense that its limit solutions correspond to solutions of
the compressible Euler and Navier--Stokes equations~\cite{Bardos:1991vf,Esposito:1994ca}, the incompressible Euler and Navier--Stokes equations~\cite{Golse:2004oe,Lions:2001sj}, the incompressible Stokes equations~\cite{Lions:2001wb} and
the incompressible Navier--Stokes--Fourier system~\cite{Levermore:2010kx}; see~\cite{Saint-Raymond2009} for an overview.
The Boltzmann equation is uniquely suited to describe flows in
the transitional molecular/continuum regime and the corresponding rarefaction effects, by virtue of its inherent characterization of deviations of the velocity distribution from local equilibrium. Applications in which rarefaction effects play a significant role are multitudinous, including gas flow problems
involving large mean free paths in high-altitude flows and hypobaric applications such as chemical vapor deposition; see \cite{Cercignani2000,Struchtrup2005} and references therein for further examples.
Moreover, the perpetual trend toward miniaturization in science and technology renders accurate descriptions of fluid flows in the transitional molecular/continuum regime of fundamental technological relevance, for instance, in nanoscale applications, micro-channel flows or flow in porous media~\cite{shen2005}.
The Boltzmann equation also provides a prototype for kinetic models in many other applications that require a description of the collective
behavior of large ensembles of small particles, for instance, in semi-conductors~\cite{jungel2009}, in plasmas and fusion and fission devices~\cite{miyamoto2004}  and
in dispersed-particle flows such as in fluidized-bed reactors~\cite{reeks1991,reeks1992,reeks1993}.

Numerical approximation of the Boltzmann equation poses a formidable challenge, on account of its high dimensional setting: for a problem in~$D$ spatial dimensions, the one-particle phase-space is~$2D$ dimensional. The corresponding computational complexity of conventional discretization methods for (integro-)differential equations, such as finite-element methods with uniform meshes, is prohibitive. Numerical approximations of the Boltzmann equation
have been predominantly based on particle methods, such as the Direct Simulation Monte Carlo (DSMC) method~\cite{Bird:1970,Bird:1994}. Convergence proofs for these methods~\cite{Wagner1992} however convey that their computational complexity depends sensitively on the Knudsen number, and the computational cost becomes prohibitive in the fluid-dynamical limit. Moreover, from an approximation perspective, DSMC can be inefficient, because it is inherent to the underlying Monte-Carlo process that the approximation error decays only as $n^{-\frac{1}{2}}$ as the number of simulation molecules,~$n$, increases; see, for instance, \cite[Thm.~5.14]{Klenke2008}. Efficient
computational modeling of fluid flows in the transitional molecular/continuum regime therefore remains an outstanding challenge.

An alternative approximation technique for the Boltzmann equation is the method of moments \cite{Grad1949,Levermore1996,Struchtrup2005}. The method of moments represents a general statistical approximation technique which identifies parameters of an approximate distribution based on its moments~\cite{Matyas1999}. Application of the method of moments to the Boltzmann equation engenders an evolution equation for the moments (weighted averages) of the phase-space distribution. An approximation based on moments is generally consistent with a restricted interest in functionals of the distribution corresponding to macroscopic properties of the fluid. The method of moments is closely related to extended thermodynamics; see, for instance,~\cite{Dreyer1987,Muller1993}.

In this paper we present a position-velocity Galerkin approximation method for the Boltzmann equation, based on a
moment-system approximation in velocity dependence and a discontinuous Galerkin approximation in position dependence.
To devise the moment-closure relation, we consider a generalization of the setting of the moment-closure problem from
Kullback--Leibler divergence~\cite{Kullback1951} to the class
of $\varphi$\nobreakdash-divergences~\cite{Csiszar1972,Abdel-Malik:2015eu}. The derived moment-closure relation engenders non-negative distributions and the corresponding moment systems are symmetric hyperbolic and tractable, in the sense that the formulation only requires the evaluation of higher-order moments of Gaussian distributions. The moment systems conserve mass, momentum and energy, and moreover dissipate an appropriate $\varphi$-divergence, analogous to the dissipation of relative entropy of the underlying Boltzmann equation, provided that the collision operator dissipates the corresponding $\varphi$-divergence relative to a suitable reference distribution. The moment systems correspond to Galerkin approximations of the Boltzmann equation in renormalized form. For the discretization in position dependence, we consider a discontinuous Galerkin finite-element method (DGFEM). We present a new numerical-flux function for the DGFEM discretization, derived from the underlying renormalized Boltzmann equation. We will show that the resulting DGFEM moment method method is entropy stable.

The remainder of this paper is organized as follows. Section~\ref{sec:BoltzProps} surveys standard structural properties
of the Boltzmann equation that have to be retained in the moment-system approximation. Section~\ref{sec:MomSys} introduces the moment-system approximation of the Boltzmann equation in velocity dependence and the corresponding moment-closure relation. In Section~\ref{sec:DGFEM}, we derive the DGFEM formulation in position dependence, and we show that the
resulting DGFEM moment method is entropy stable. Section~\ref{sec:NumRes} illustrates the properties of the proposed DGFEM moment method for a shock tube Riemann problem \cite{Courant1999,Toro2013}. Finally, Section~\ref{sec:Conc} presents a concluding discussion.

\section{Properties of the Boltzmann equation}
\label{sec:BoltzProps}
Consider a monatomic gas contained in a fixed spatial domain $\Omega\subset\mathbb{R}^D $. Kinetic theory describes the state of such a gas by a non-negative (phase-space) density $f=f(t,\bm x,\bm v)$ over the single-particle phase space $\Omega\times\mathbb{R}^D$. The evolution of $f$ is considered to be governed by the Boltzmann equation,
\begin{align}\label{eq:Boltzmann}
\partial_t f + v_i\partial_{x_i} f=\CC(f)
\end{align}
where the summation convention applies to repeated indices and the collision operator $f\mapsto\CC(f)$ acts only on the $\bm v=(v_1,\ldots,v_D)$ dependence  of $f$ locally at each $(t,\bm x)$.
The collision operator is assumed to possess certain conservation, symmetry and dissipation properties, viz.,
conservation of mass, momentum and energy, invariance under Galilean transformations and dissipation of appropriate entropy functionals.
Moreover, it is assumed that the collision operator exhibits certain positivity properties. These fundamental properties of the collision operator are treated in further detail below. Our treatment of the conservation and symmetry properties is standard (see, for instance,~\cite{Levermore1996}) and is presented merely for coherence and completeness. For the entropy-dissipation property, we consider a generalization of the usual (relative) entropy to $\varphi$-divergences~\cite{Csiszar1972,Abdel-Malik:2015eu} for collision operators $f\mapsto\CC(f)$ that are non-negative outside the support of~$f$, to enable an exploration of the moment-closure problem in an extended setting.

To elaborate the conservation properties of the collision operator,
let $\langle \cdot \rangle$ denote integration in the velocity dependence of any scalar, vector or matrix valued measurable function
over $D$\nobreakdash-dimensional Lebesgue measure. A function $\alpha:\IR^D\to\IR$ is called a {\em collision invariant\/} of $\CC$
if
\begin{align}
\label{eq:DefCollInvariant}
\langle \alpha\,\CC(f) \rangle = 0  \qquad \forall f\in \Domain(\CC),
\end{align}
where $\Domain(\CC)\subset{}L^1(\IR^D,\IR_{\geq{}0})$ denotes the domain of~$\CC$, which we consider to be a subset of the almost everywhere nonnegative
Lebesgue integrable functions on~$\IR^D$.
Equation~\EQ{Boltzmann} associates a scalar conservation law with each collision invariant:
\begin{equation}
\label{eq:ConsvLaw}
\partial_t\langle \alpha f\rangle+\partial_{x_i} \langle v_i \alpha f \rangle = 0
\end{equation}
We insist that $\{1,v_1,\ldots,v_D,|\bm v|^2\}$ are collision invariants of $\CC$ and that
the span of this set contains all collision invariants, i.e.
\begin{equation}\label{eq:CollInvariant}
\langle{}\alpha\,\CC(f)\rangle=0\quad\forall{}f\in\Domain(\CC)
\quad\Leftrightarrow\quad
\alpha\in\mathrm{span}\{1,v_1,\ldots,v_D,|\bm v|^2\}=:\IE.
\end{equation}
The moments $\langle{}f\rangle$,
$\langle{}v_if\rangle$ and $\langle{}|{\bm v}|^2f\rangle$, correspond to mass-density, the (components of) momentum-density and energy-density, respectively.
Accordingly, the conservation law~\EQ{ConsvLaw} implies that~\EQ{Boltzmann} conserves mass, momentum and energy.

The assumed symmetry properties of the collision operator pertain to commutation with translational and rotational transformations.
In particular, for all vectors $\bm u\in\mathbb{R}^D$ and all orthogonal tensors $\mathcal{O}:\IR^D\to\IR^D$, we define the translation transformation
$\mathcal{T}_{\bm u}:\Domain(\CC)\to\Domain(\CC)$ and the rotation transformation $\mathcal{T}_{\mathcal{O}}:\Domain(\CC)\to\Domain(\CC)$ by:
\begin{alignat}{2} \label{eq:Tran}
(\mathcal{T}_{\bm u}f)(\bm v)&=f(\bm u-\bm v)&\qquad&\forall{}f\in\Domain(\CC)
\\
\label{eq:Rot}
(\mathcal{T}_{\mathcal{O}}f)(\bm v)&=f(\mathcal{O}^*\bm v) &\qquad&\forall{}f\in\Domain(\CC)
\end{alignat}
with $\mathcal{O}^*$ the Euclidean adjoint of $\mathcal{O}$. Note that the above transformations act on the $\bm v$-dependence only.
It is assumed that $\CC$ possesses the following symmetries:
\begin{equation}
\label{eq:GalilInvar}
\CC(\mathcal{T}_{\bm u}f)=\mathcal{T}_{\bm u}\CC(f),\qquad
\CC(\mathcal{T}_{\mathcal{O}}f)=\mathcal{T}_{\mathcal{O}}\CC(f)
\end{equation}
The symmetries (\ref{eq:GalilInvar}) imply that (\ref{eq:Boltzmann}) complies with Galilean invariance, i.e.
if $f(t,\bm x,\bm v)$ satisfies the Boltzmann equation~\EQ{Boltzmann}, then for arbitrary ${\bm u}\in\IR^D$ and arbitrary orthogonal $\mathcal{O}:\IR^D\to\IR^D$, so
do $f(t,\bm x - {\bm u}t,{\bm v}-{\bm u})$ and $f(t,\mathcal{O}^*{\bm x},\mathcal{O}^*{\bm v})$.

The entropy dissipation property of~$\CC$ is considered in the extended setting of~\cite[Sec.~7]{Levermore1996}, from which we
derive the following definition: a convex function $\HH:\IR_{\gneq{}0}\to\IR$ is called an {\em entropy for $\CC$\/} if
\begin{equation}
\label{eq:Dissipation}
\langle \HH'(f)\,\CC(f) \rangle \leq 0, \qquad \forall f\in\Domain(\CC)
\end{equation}
with $\HH'(f)$ the derivative of $\HH(f)$, and if for every $f\in\Domain(\CC)$ the following equivalences hold:
\begin{equation}
\label{eq:Equilibrium}
\CC(f) = 0
\quad\Leftrightarrow\quad
\langle \HH'(f)\,\CC(f)\rangle  = 0
\quad\Leftrightarrow\quad
\HH'(f)\in\IE
\end{equation}
Relation (\ref{eq:Dissipation}) implies that~$\CC$ dissipates the local entropy density $\langle\HH(\cdot)\rangle$, which leads to an
abstraction of Boltzmann's H-theorem for~\EQ{Boltzmann}, asserting that solutions of the Boltzmann equation (\ref{eq:Boltzmann}) satisfy the
local entropy-dissipation law:
\begin{align}
\label{eq:EntDiss}
 \partial_t\langle \HH(f) \rangle+\partial_{x_i} \langle v_i \HH(f) \rangle = \langle \CC(f)\, \HH'(f)\rangle \leq 0\,.
\end{align}
The functions $\langle \HH(f) \rangle$,  $\langle v_i \HH(f) \rangle$ and $\langle \HH'(f)\, \CC(f)\rangle$ are referred to as entropy density, entropy flux and entropy-dissipation rate, respectively.
The first equivalence in~(\ref{eq:Equilibrium}) characterizes local equilibria of~$\CC$
by vanishing entropy dissipation, while the second equivalence indicates the form of such local equilibria.
For spatially homogeneous initial data, $f_0$, equations~\EQ{Dissipation} and~\EQ{Equilibrium} suggest that equilibrium solutions, $\feq$,
of~\EQ{Boltzmann} are determined by:
\begin{equation}
\label{eq:LegEq}
\feq=\argmin\big\{\langle\HH(f)\rangle:f\in\Domain(\CC),\langle\alpha f \rangle=\langle{}\alpha f_0\rangle\:\:\forall\alpha\in\IE\},
\end{equation}
Equation~\EQ{LegEq} identifies equilibria as minimizers\footnote{We adopt the sign convention of diminishing entropy.} of the entropy, subject to the constraint that the invariant moments are identical to the invariant moments of the initial distribution.

The standard definition of entropy corresponds to a density $f\mapsto{}\langle{}f\log{}f+f\alpha\rangle{}$ where $\alpha\in\IE$ is any collision invariant. The corresponding local equilibria of $\CC(f)$ defined by (\ref{eq:Equilibrium}) are characterized by Maxwellians $\MM$, i.e. distributions of the form
\begin{equation} \label{eq:Maxwellian}
\MM({\bm v}):=
\MM_{(\varrho,{\bm u},T)}({\bm v}) :=
\frac{\varrho}{(2\pi{}RT)^{\frac{D}{2}}}\exp\left(-\frac{|\bm v-\bm u|^2}{2RT}\right)
\end{equation}
for some $(\varrho, {\bm u}, T)\in\mathbb{R}_{>0}\times\mathbb{R}^D\times\mathbb{R}_{>0}$ and a certain gas constant $R\in\IR_{>0}$. It is noteworthy that $\log\MM\in\IE$ and, therefore, the Kullback-Leibler divergence~$\langle{}f\log{}(f/\MM)\rangle$ of~$f$ with  respect to~$\MM$ is equivalent to $\langle{}f\log{}f\rangle$ in the sense of dissipation characteristics.

In this work, we depart from the standard relative entropy for~\EQ{Boltzmann} corresponding to the Kullback-Leibler divergence.
Instead, we consider entropies based on
particular $\varphi$\nobreakdash-divergences~\cite{Ali1966,Csiszar1972,Abdel-Malik:2015eu}.
These $\varphi$\nobreakdash-divergences generally preclude the usual physical and information-theoretical
interpretations of relative entropy, but still provide a meaningful entropy density in accordance with~(\ref{eq:Dissipation}) and~(\ref{eq:Equilibrium}).
The considered $\varphi$\nobreakdash-divergences yield a setting in which entropy-minimization based moment-closure approximations to~\EQ{Boltzmann}
are not impaired by non-realizability, exhibit bounded fluxes in the vicinity of equilibrium, and are numerically tractable.

We will admit distributions that vanish on sets with nonzero measure. To
accommodate such distributions, we introduce
an auxiliary non-negativity condition on the collision operator, in addition
to~(\ref{eq:Dissipation}) and~(\ref{eq:Equilibrium}). The non-negativity
condition insists that $\CC(f)$ cannot be negative on zero sets of~$f$:
\begin{equation}
\label{eq:NonNeg}
\CC(f)\big|_{\mathrm{supp}^c(f)}\geq{}0
\end{equation}
where $\mathrm{supp}^c(f)$ denotes zero set of~$f$, i.e. the complement in $\IR^D$ of the closed support of~$f$.
Condition~(\ref{eq:NonNeg}) encodes that the collision operator cannot create
locally negative distributions. It can be
verified that~(\ref{eq:NonNeg}) holds for a wide range of collision operators,
including the
BGK operator~\cite{Bhatnagar:1954hc}, the multi-scale generalization of the BGK
operator introduced in~\cite{Levermore1996},
and all collision operators that are characterized by a (non-negative) collision
kernel.

\begin{remark}
\label{rem:remark1}
The adoption of $\varphi$-divergence-based entropies insists that such entropies
satisfy~\EQ{Dissipation} and~\EQ{Equilibrium} for a meaningful class of collision
operators in compliance with~(\ref{eq:NonNeg}). In~\cite{Abdel-Malik:2015eu} it is
shown that the class of admissible collision operators includes the BGK
operator~\cite{Bhatnagar:1954hc} and the multi-scale generalization of the BGK
operator in~\cite{Levermore1996}.
\end{remark}

\section{Moment-system approximation of the Boltzmann equation}
\label{sec:MomSys}
Our semi-discretization of the Boltzmann equation
with respect to the velocity dependence is based on velocity moments
of the one-particle marginal. An inherent aspect of considering a finite number of
moment equations derived from~\EQ{Boltzmann} is that low-order moments are
generally coupled to higher-order ones. Consequently, a closed set of
equations for the moments cannot be readily formulated, and a closure
relation is required. Closed moment systems can generally be conceived of
as Galerkin-subspace approximations of~(\ref{eq:Boltzmann}) in renormalized form.

To derive the moment equations from~\EQ{Boltzmann},
let $\IM$ denote a finite-dimensional subspace of $D$-variate polynomials and
let $\{m_i(\bm v)\}_{i=1}^M$  with $M=\dim\IM$ represent a corresponding basis. Denoting the
column $M$\nobreakdash-vector of these
basis elements by $\bm m$, it holds that the moments
$\langle{}\bm m{}f\rangle$ of the one-particle marginal satisfy:
\begin{align} \label{eq:MomSys}
\partial_t\langle \bm m  f\rangle+\partial_{x_i}\langle v_i\bm m  f\rangle =
\langle \bm m\CC(f)\rangle
\end{align}
provided that~$f\in
\IF:=\big\{f\in{}\Domain(\CC): f\geq0,\, mf\in{L}^1(\mathbb{R}^{D}),\,\bm v
mf\in{L}^1(\mathbb{R}^{D},\IR^D),\,m\CC(f)\in{}L^1(\IR^D) \ \forall
m\in\IM\big\}$
almost everywhere in the considered time interval $(0,T)$ and the spatial domain
$\Omega$.
The provision~$f\in\IF$ has been confirmed in specific settings of~\EQ{Boltzmann} but
not for the general case; see~\cite[Sec. 4]{Levermore1996}.
The moment system~\EQ{MomSys} constitutes~$M$ relations between
$(2+D)$ $\IR^M$-valued functions, viz.
the density $\langle{}\bm m f\rangle$, the fluxes
$\langle{}v_i \bm m f\rangle$ and the production term $\langle \bm m\CC(f)\rangle$,
and is therefore not closed.
Moment systems are generally closed by constructing an approximation to the
distribution function from the densities and
then evaluating the fluxes and production terms for the approximate
distribution. Denoting by
$\IA\subseteq\IR^M$ a suitable class of moments, a function $\ff:\IA\to\IF$ must
be
specified such that~$\ff$ realizes the moments in~$\IA$, i.e. $\langle\bm
m\ff(\bm\mu)\rangle=\bm\mu$ for all $\bm\mu\in\IA$,
and if~$f$ satisfies~\EQ{Boltzmann} then $\ff(\langle{}\bm m f\rangle)$ constitutes
a suitable (in a sense to be made more precise below) approximation
to~$f$.
Approximating the moments in~\EQ{MomSys} by $\bm\mu\approx\langle\bm m
f\rangle$ and replacing $f$ in~\EQ{MomSys} by the
approximation $\ff(\bm\mu)$, one obtains the following closed system for the
approximate moments:
\begin{equation}
\label{eq:ClsMomSys}
\partial_t\bm\mu+\partial_{x_i}\langle v_i\bm m
\mathcal{F}(\bm\mu)\rangle=\langle\bm m \CC(\mathcal{F}(\bm\mu))\rangle.
\end{equation}
The closed moment system~\EQ{ClsMomSys} is essentially defined by the polynomial
subspace, $\IM$, and the closure relation,~$\ff$.
A subspace/closure-relation pair $(\IM,\ff)$ is appropriate if the corresponding
moment system~(\ref{eq:ClsMomSys}) is well posed and retains the fundamental
structural properties of the Boltzmann equation:
conservation of mass, momentum and energy, Galilean invariance and dissipation
of an entropy functional. Further conditions
may be considered, e.g. that the fluxes and production terms can
be efficiently evaluated by means of numerical quadrature.
The conservation properties and Galilean invariance of~\EQ{Boltzmann} can generally
be transferred to~\EQ{ClsMomSys} by a suitable selection of the subspace~$\IM$,
namely that $\IM$ contains the collision invariants $\IE$, and is closed under the
actions of $\mathcal{T}_{\bm u}$ and $\mathcal{T}_{\mathcal{O}}$ (cf.~\EQ{Tran} and~\EQ{Rot}).
Entropy dissipation must be ensured by the closure relation, $\ff$.

We consider a moment-closure relation deriving from minimization of a $\varphi$\nobreakdash-divergence
relative to a suitable background measure, subject to the moment constraints; see also~\cite{Abdel-Malik:2015eu}.
An example of the class of $\varphi$\nobreakdash-divergences that we envisage is provided by:
\begin{equation}
\label{eq:varphiN}
\varphi(s)=s\bigg(\frac{N^2}{N+1}s^{1/N}-N\bigg)+\frac{N}{N+1}
\end{equation}
with $N\in\mathbb{N}$. We assume that the considered $\varphi$\nobreakdash-divergence is strictly convex and
represents an entropy
for the collision operator. We denote by $\BB$ a strictly positive background distribution. We assume that $\BB$ is
independent of $(t,\bm x)$. However, this assumption can be moderated. We consider the closure relation $\ff:\IA\to\IF$ according~to:
\begin{equation}
\label{eq:EntMin}
\ff(\bm\mu) =
\underset{f\in\IF}{\argmin}\big\{\big\langle\BB\,\varphi(f/\BB)\big\rangle:
f\in\Domain(\CC),\langle \bm m f\rangle=\bm
\mu\big\}
\end{equation}
Assuming that~\EQ{EntMin} admits a solution (see~\cite{Abdel-Malik:2015eu}),
this solution can be characterized as a stationary point of the Lagrangian
$(f,\bm\lambda)\mapsto\langle{}\BB\,\varphi(f/\BB){}
\rangle+\bm\lambda\cdot(\bm\mu-\langle { } \bm m f\rangle)$, where $\bm\lambda$ represents a
Lagrange multiplier. The stationarity condition implies
$\varphi'(f/\BB)-\bm\lambda\cdot\bm m=0$ and, hence, the moment-closure relation:
\begin{equation}
\label{eq:Cls}
\ff(\bm\mu)
= \BB\,\psi(\bm\lambda\cdot\bm m)
\end{equation}
with $\psi$ a suitable right inverse of~$\varphi'$. In particular,
we define
\begin{equation}
\label{eq:psi}
\psi(s)=
\begin{cases}
(\varphi')^{-1}(s)&\quad\text{if }s>\varphi'(0)
\\
0&\quad\text{if }s\leq\varphi'(0)
\end{cases}
\end{equation}
By virtue of the strict convexity of $\varphi$, the derivative $\varphi'$ is an increasing function
and the inverse in~\EQ{psi} is uniquely defined.
It is important to note that $\varphi'$ corresponds to a left inverse
of~$\psi$ on the support of~$\ff(\bm\mu)$, i.e. on the closure of
$\{\bm v\in\IR^D:(\varphi')^{-1}(\bm\lambda\cdot\bm m(\bm v))>0\}$:
\begin{equation}
\varphi'\big(\psi(\bm\lambda\cdot\bm m(\bm v))\big)=
\bm\lambda\cdot\bm m(\bm v)
\qquad\text{for all }\bm{v}\in
\mathrm{cl}\big(\{\bm v\in\IR^D:(\varphi')^{-1}(\bm\lambda\cdot\bm m(\bm v))>0\}\big)
\end{equation}
For all $\bm v\in\mathrm{supp}^c(\ff(\bm\mu))$, it holds that
that $\psi(\bm\lambda\cdot\bm m(\bm v))=0$ but
generally $\varphi'(0)\neq\bm\lambda\cdot\bm{m}(\bm v)$.
To establish that under suitable boundary conditions, the moment system~\EQ{ClsMomSys} with
closure relation~\EQ{Cls} dissipates the
relative entropy with density $f\mapsto\langle\BB\,\varphi(f/\BB)\rangle$, we
will show that if~$\ff:=\ff(\bm{\mu})$ satisfies~\EQ{ClsMomSys} then
\begin{equation}
\label{eq:dissipationineq}
\frac{d}{dt}\int_{\Omega}\big\langle\BB\,\varphi(\ff/\BB)\big\rangle
+
\int_{\partial\Omega}
\big\langle v_n\BB\,\varphi(\ff/\BB)\big\rangle
\leq{}0
\end{equation}
where $v_n=\bm{v}\cdot\bm{n}$ with $\bm{n}$ the exterior unit normal vector on~$\partial\Omega$.
To this end, we first note that by the chain rule, we have:
%
%
\begin{equation}
\label{eq:=T1+T2}
\frac{d}{dt}\int_{\Omega}\big\langle\BB\,\varphi(\ff/\BB)\big\rangle
+
\int_{\partial\Omega}
\big\langle v_n\BB\,\varphi(\ff/\BB)\big\rangle
=
\int_{\Omega}
\big(
\partial_t
\big\langle\BB\,\varphi(\ff/\BB)\big\rangle
+
\partial_{x_i}\big\langle v_i\BB\,\varphi(\ff/\BB)\big\rangle\big)
=
\int_{\Omega}
(T_1+T_2)
\end{equation}
with
\begin{align*}
T_1&=
\big\langle\varphi'(\ff/\BB)\big(\partial_t\ff
+
v_i\partial_{x_i}\ff\big)\big\rangle
\\
T_2&=
\big\langle \big(\varphi(\mathcal{F}/\BB) - (\ff/\BB)\varphi'(\ff/\BB)\big)\big(\partial_t\BB+v_i\partial_{x_i}\BB\big)
\big\rangle
\end{align*}
On account of $\varphi'(\ff/\BB)=\bm\lambda\cdot\bm{m}$ on $\mathrm{supp}(\ff)$ and~\EQ{ClsMomSys},
we obtain for the first term:
\begin{equation}
\label{eq:T1}
\begin{aligned}
T_1&=
\big\langle(\bm\lambda\cdot\bm{m})(\partial_t\ff+v_i\partial_{x_i}\ff)\big\rangle
+
\int_{\mathrm{supp}^c(\ff)}
(\varphi'(0)-\bm\lambda\cdot\bm{m})(\partial_t\ff+v_i\partial_{x_i}\ff)
\\
&=\big\langle(\bm\lambda\cdot\bm{m})\,\CC(\ff)\big\rangle
+
\int_{\mathrm{supp}^c(\ff)}
(\varphi'(0)-\bm\lambda\cdot\bm{m})(\partial_t\ff+v_i\partial_{x_i}\ff)
\\
&=\big\langle\varphi'(\ff/\BB)\,\CC(\ff)\big\rangle
+
\int_{\mathrm{supp}^c(\ff)}
(\varphi'(0)-\bm\lambda\cdot\bm{m})(\partial_t\ff+v_i\partial_{x_i}\ff-\CC(\ff))
\end{aligned}
\end{equation}
The first term in the ultimate expression in~\EQ{T1} is non-positive under the standing assumption that
$f\mapsto\langle\BB\varphi(f/\BB)\rangle$ represents an entropy for the collision operator
in accordance with~\EQ{Dissipation}. Moreover, on the zero set $\mathrm{supp}^c(\ff)$ it holds that
$\varphi'(0)\geq\bm\lambda\cdot\bm{m}(\bm{v})$, $\partial_t\ff+\partial_{x_i}\ff=0$, and $\CC(\ff)\geq{}0$.
The latter inequality follows directly from the non-negativity condition~\EQ{NonNeg}. Hence, $T_1$ is
non-positive. The second term, $T_2$, in the right member of~\EQ{=T1+T2} vanishes because the background distribution~$\BB$ is independent of~$(t,\bm{x})$.

It is noteworthy that the entropy-dissipation inequality~\EQ{dissipationineq}
can be extended to non-uniform background distributions.
The entropy-dissipation inequality holds, for instance, if the background distribution is a locally Maxwellian
flow~\cite[Appendix~2]{Grad1949} or a solution to the Vlasov equation, in which case $\partial_t\BB+v_i\partial_{x_i}\BB=0$.
Let us also allude to the fact that the analysis in~\EQ{=T1+T2}-\EQ{T1} relies on continuity of $\BB\varphi(\ff/\BB)$
in $(\bm x,\bm v)$, for otherwise the chain rule in~\EQ{=T1+T2} is invalid.

To demonstrate that the moment system~\EQ{ClsMomSys} with closure relation~\EQ{Cls}
corresponds to a symmetric hyperbolic system, we first reformulate~\EQ{ClsMomSys} in terms
of the Lagrange multipliers. The constraints in~\EQ{Cls} imply that
$\bm\mu=\langle\bm{m}\BB\psi(\bm\lambda\cdot\bm{m})\rangle$. Hence, we have
\begin{equation}
\label{eq:hypsysmom}
\partial_t\bm{\mu}
=
\partial_t\langle\bm{m}\BB\psi(\bm\lambda\cdot\bm{m})\rangle
=
\langle\bm{m}\psi(\bm\lambda\cdot\bm{m})\partial_t\BB\rangle+\bm{A}_0(\bm{\lambda})\partial_t\bm{\lambda}
\end{equation}
with $\bm{A}_0(\bm{\lambda})=\langle\bm{m}\otimes\bm{m}\,\BB\psi'(\bm{\lambda}\cdot\bm{m})\rangle$. For the flux terms, we obtain
\begin{equation}
\label{eq:hypsysflux}
\partial_{x_i}\langle{}v_i\bm{m}\BB\psi(\bm{\lambda}\cdot\bm{m})\rangle
=
\langle\bm{m}\psi(\bm\lambda\cdot\bm{m})v_i\partial_{x_i}\BB\rangle
+
\bm{A}_i(\bm{\lambda})\partial_{x_i}\bm{\lambda}
\end{equation}
with $\bm{A}_i(\bm{\lambda})=\langle{}v_i\bm{m}\otimes\bm{m}\,\BB\psi'(\bm{\lambda}\cdot\bm{m})\rangle$.
By virtue of~\EQ{hypsysmom} and~\EQ{hypsysflux}, the moment system can be recast as
the following quasi-linear system for the Lagrange multipliers:
\begin{equation}
\label{eq:HypSys}
{\bm A}_0(\bm\lambda)\frac{\partial{\bm\lambda}}{\partial{}t}
+
\sum_{i=1}^D{\bm A}_i(\bm\lambda)\frac{\partial\bm\lambda}{\partial{}x_i}
=
{\bm s}(\bm\lambda)
\end{equation}
with
\begin{equation}
\label{eq:s}
\bm{s}(\bm{\lambda})=
\big\langle \bm m \CC(\BB\psi(\bm\lambda\cdot\bm m))\big\rangle
-
\big\langle\bm{m}\psi(\bm{\lambda}\cdot\bm{m})(\partial_t{\BB}+v_i\partial_{x_i}\BB)\rangle
\end{equation}
System~\EQ{HypSys} is symmetric hyperbolic if~${\bm A}_0,{\bm A}_1,\ldots,{\bm A}_D$ are symmetric and
${\bm A}_0$ is positive definite.
The symmetry of~${\bm A}_0,{\bm A}_1,\ldots,{\bm A}_D$ is evident. To corroborate the positive definiteness of $\bm{A}_0$,
we note that for any $M$-vector $\bm{\theta}$ there holds
\begin{equation}
\label{eq:PosDef}
\bm{\theta}\cdot\bm{A}_0(\bm{\lambda})\bm{\theta}=
\big\langle(\bm{\theta}\cdot\bm{m})^2\BB\psi'(\bm{\lambda}\cdot\bm{m})\big\rangle\geq{}0
\end{equation}
The inequality holds because each of the factors
$(\bm{\theta}\cdot\bm{m})^2$, $\BB$ and $\psi'(\bm{\lambda}\cdot\bm{m})$ is non-negative.
For $\bm{\theta}\neq{}0$, the inequality in~\EQ{PosDef} is strict, because the roots of the
polynomial $\bm{\theta}\cdot\bm{m}(\bm{v})$ are confined to a set of measure zero, $\BB$ is
strictly positive by assumption, and $\psi'$ is strictly positive on $(\varphi'(0),\infty)$.
The matrix~$\bm{A}_0(\bm{\lambda})$ is therefore indeed positive definite.
By virtue of its quasi-linear symmetric hyperbolicity, the system~\EQ{HypSys} is linearly well posed.
Moreover, under suitable conditions on the initial data, local-in-time
existence of solutions can be established~\cite{Majda:1984wj}.
It is to be noted that the term corresponding to $\partial_t\BB+v_i\partial_{x_i}\BB$ in the production
term according to~\EQ{s} can cause blow up of solutions to the hyperbolic system~\EQ{HypSys}
in the limit $t\rightarrow\infty$. Hence, the hyperbolic character of~\EQ{ClsMomSys} with closure
relation~(\ref{eq:Cls}) ensures stability of solutions only in finite time.
If the background distribution $\BB$ is selected such that $\partial_t\BB+v_i\partial_{x_i}\BB$ vanishes,
then the production term exhibits the usual dissipation properties corresponding to the collision operator,
and the stability provided by the symmetric-hyperbolic character of the equations extends to the ad-infinitum limit.

The moment system~\EQ{ClsMomSys} can alternatively be construed as a Galerkin subspace approximation of the
Boltzmann equation in renormalized form; see also~\cite{Abdel-Malik:2015eu}.
This Galerkin-approximation interpretation can for instance
prove useful in constructing error estimates for~\EQ{ClsMomSys} and in deriving
structural properties. To elucidate the Galerkin form of~\EQ{ClsMomSys},
we define the renormalization map $\beta:\IM\to\IF$ according to $\beta(g)=\BB\psi(g)$ and
we observe that for all $\bm{\mu}\in\IR^M$ there exists a corresponding element $g\in\IM$ such
that $\ff(\bm{\mu})=\beta(g)$. In particular, $g=\bm{\lambda}\cdot\bm{m}$ with~$\bm{\lambda}$
the Lagrange multipliers associated with the constrained minimization problem~\EQ{EntMin}.
The moment system~\EQ{ClsMomSys} can then be recast into the Galerkin form:
\begin{multline}  \label{eq:GalLevMomCls}
  \text{\it Find }
  g\in{}\mathscr{L}\big((0,T)\times\Omega;\IM\big):\\
  \partial_t\big\langle m \beta(g)\big\rangle+\partial_{x_i}\big\langle
mv_i\beta(g)\big\rangle = \big\langle m\CC(\beta{}(g))\big\rangle
  \quad
  \forall m\in{}\IM\text{ a.e. }(t,\bm x)\in(0,T)\times\Omega
\end{multline}
where $\mathscr{L}\big((0,T)\times\Omega;\IM\big)$ represents a suitable vector space
of functions from~$(0,T)\times\Omega$ into~$\IM$.

\section{Spatial discontinuous Galerkin finite-element approximation}
\label{sec:DGFEM}
For the discretization of~\EQ{ClsMomSys} with respect to the position dependence, we consider
the discontinuous Galerkin finite-element method~\cite{Di-Pietro:2012kx}.
Let $\mathcal{H}:=\{h_1,h_2,\ldots\}\subset\IR_{>0}$ denote
a strictly decreasing sequence of mesh parameters
whose only accumulation point is~$0$. Consider a corresponding mesh sequence $\mathcal{T}^{\mathcal{H}}$,
viz., a sequence of covers of the domain by non-overlapping element domains $\kappa\subset\Omega$. We impose
on $\mathcal{T}_{\mathcal{H}}$ the standard conditions of regularity, shape-regularity and quasi-uniformity
with respect to~$\mathcal{H}$; see, for instance, \cite{Di-Pietro:2012kx} for further details. For any $h\in\mathcal{H}$,
we indicate by $V^{h,p}(\Omega)$ the DG finite-element approximation space of discontinuous element-wise $D$-variate polynomials
of degree $\leq{}p$:
\begin{equation}
V^{h,p}(\Omega)=\{v\in{}L^2(\Omega):v|_{\kappa}(\bm{x})\in\mathrm{span}\{x_1^{p_1}x_2^{p_2}\cdots{}x_D^{p_D}\},p_1+\cdots+p_D\leq{}p\}
\end{equation}
The $p$-dependence of $V^{h,p}$ is contextual and will generally be suppressed. We denote by
\begin{equation}
V^h(\Omega,\IM)=V^h(\Omega)\times\IM=\{\lambda_1m_1+\cdots+\lambda_Mm_M:\lambda_i\in{}V^h(\Omega)\}
\end{equation}
the extension of $V^h$ to $\IM$-valued functions.

To facilitate the presentation of the DGFEM formulation, we introduce some further notational conventions. For any $h\in\mathcal{H}$, we indicate by $\mathcal{I}^h=\{\mathrm{int}(\partial\kappa\cap\partial\hat{\kappa}):\kappa,\hat{\kappa}\in\mathcal{T}^h,\kappa\neq\hat{\kappa}\}$ the collection of inter-element edges, by $\mathcal{B}^h=\{\mathrm{int}(\partial\kappa\cap\partial\Omega):\kappa\in\mathcal{T}^h\}$ the collection of boundary edges and by $\mathcal{S}^h=\mathcal{B}^h\cup\mathcal{I}^h$ their union.
With every edge we associate a unit normal vector~$\bm{\nu}_e$. The orientation of~$\bm{\nu}_e$ is arbitrary
except on boundary edges where $\bm{\nu}_e=\bm{n}|_e$. For all interior edges, let $\kappa_{\pm}^e\in\mathcal{T}^h$ be
the two elements adjacent to the edge~$e$ such that the orientation of $\bm{\nu}_e$ is exterior to~$\kappa_+$.
We define jump $\jump{\cdot}_{e}$ and mean $\mean{\cdot}_{e}$ operators according to:
\begin{equation}
\label{eq:jumpmean}
\jump{v}_e=
\begin{cases}
(v_+-v_-)&\text{ if }e\in\mathcal{I}^h
\\
v_+&\text{ if }e\in\mathcal{B}^h
\end{cases}
\qquad
\mean{v}_e=
\begin{cases}
(v_++v_-)/2&\text{ if }e\in\mathcal{I}^h
\\
v_+&\text{ if }e\in\mathcal{B}^h
\end{cases}
\end{equation}
where $v_+$ and $v_-$ refer to the restriction of the traces of $v|_{\kappa_+}$ and $v|_{\kappa_-}$ to~$e$.
To derive the DG formulation of the closed moment system~\EQ{ClsMomSys}, we note that for any $\bm{w}\in{}[V^{h}(\Omega)]^M$ there holds:
\begin{equation}
\label{eq:DGweight}
\sum_{\kappa\in\mathcal{T}^h}\int_{\kappa}\bm{w}\cdot\partial_t\bm\mu
+
\sum_{\kappa\in\mathcal{T}^h}\int_{\kappa}
\bm{w}\cdot\partial_{x_i}\langle v_i\bm m
\mathcal{F}(\bm\mu)\rangle
=
\sum_{\kappa\in\mathcal{T}^h}\int_{\kappa}
\bm{w}\cdot
\langle\bm m \CC(\mathcal{F}(\bm\mu))\rangle.
\end{equation}
The second term in the left member of~\EQ{DGweight} can be recast into
\begin{multline}
\label{eq:DGform2}
\sum_{\kappa\in\mathcal{T}^h}\int_{\kappa}
\bm{w}\cdot\partial_{x_i}\langle v_i\bm m
\mathcal{F}\rangle
=
\sum_{\kappa\in\mathcal{T}^h}
\int_{\partial\kappa}
\bm{w}\cdot
\langle v_i\bm{m}
\mathcal{F}\rangle\nu_i^{\kappa}
-
\sum_{\kappa\in\mathcal{T}^h}\int_{\kappa}
\partial_{x_i}\bm{w}\cdot\langle v_i\bm m
\mathcal{F}\rangle
\\
=
\sum_{e\in\mathcal{S}^h}
\int_e
\jump{\bm{w}}\cdot
\langle v_{\nu}\bm{m}
\mean{\mathcal{F}}\rangle
+
\sum_{e\in\mathcal{I}^h}
\int_e
\mean{\bm{w}}\cdot
\jump{\langle v_{\nu}\bm{m}
\mathcal{F}\rangle}
-
\sum_{\kappa\in\mathcal{T}^h}\int_{\kappa}
\partial_{x_i}\bm{w}\cdot\langle v_i\bm m
\mathcal{F}\rangle
\end{multline}
with $\bm{\nu}^{\kappa}$ the exterior unit normal vector to $\partial\kappa$ and~$v_{\nu}=\bm{v}\cdot\bm{\nu}$.
The first identity in~\EQ{DGform2} follows from the product rule and integration by parts.
The second identity results from a rearrangement of terms. Implicit in the identities in~\EQ{DGform2} is
the assumption that~$\ff$ is sufficiently smooth within the elements to permit integration by parts and
define traces on $\partial\kappa$. If, moreover, $\langle{}v_{\nu}\bm{m}\ff\rangle$ is continuous
across the interior edges, then the second term in the right member of~\EQ{DGform2} can be removed, and $\mean{\ff(\bm{\mu})}$
in the first term can be replaced by any $\hat{\ff}(\bm{\mu}_+,\bm{\mu}_-)$ subject to the consistency condition
$\hat{\ff}(\bm{\mu},\bm{\mu})=\ff(\bm{\mu})$. On boundary edges, the external moment
vector $\bm{\mu}_-$ is to be conceived of as exogenous data in accordance with boundary conditions.
It then holds that
\begin{equation}
\label{eq:DGform1}
[\partial_t\bm{\mu},\bm{w}]
+
a(\bm{\mu};\bm{w})
=
s(\bm{\mu};\bm{w})
\end{equation}
with
\begin{align}
[\partial_t\bm{\mu},\bm{w}]&=
\sum_{\kappa\in\mathcal{T}^h}\int_{\kappa}
\bm{w}\cdot\partial_t\bm\mu
\\
a(\bm{\mu};\bm{w})
&=\sum_{e\in\mathcal{S}^h}
\int_{e}
\jump{\bm{w}}\cdot
\big\langle v_{\nu}\bm{m}
\hat{\ff}(\bm\mu_+,\bm\mu_-)\big\rangle
-
\sum_{\kappa\in\mathcal{T}^h}\int_{\kappa}
\partial_{x_i}\bm{w}\cdot\big\langle v_i\bm m
\mathcal{F}\big\rangle
\label{eq:aDG}
\\
s(\bm{\mu};\bm{w})
&=
\sum_{\kappa\in\mathcal{T}^h}\int_{\kappa}
\bm{w}\cdot
\big\langle\bm m \CC(\mathcal{F}(\bm\mu))\big\rangle
\end{align}
is consistent with~\EQ{ClsMomSys} in the sense that any solution to~\EQ{ClsMomSys} that is sufficiently regular
in the aforementioned sense satisfies~\EQ{DGform1} for all $\bm{w}\in[V^h(\Omega)]^M$. The DGFEM approximation
of~\EQ{ClsMomSys} is obtained by replacing $\bm\mu$ in~\EQ{DGform1} by an approximation $\bm\mu{}^h$
in $\mathscr{L}(0,T;[V^h(\Omega)]^M)$ according to:
\begin{equation}
\label{eq:DGform}
\text{\em Find }\bm\mu{}^h\in{}\mathscr{L}(0,T;[V^h(\Omega)]^M):
\qquad
[\partial_t\bm\mu{}^h,\bm{w}]
+
a(\bm{\mu}^h;\bm{w})
=
s(\bm{\mu}^h;\bm{w})
\qquad\forall\bm{w}\in{}[V^h(\Omega)]^M
\end{equation}
a.e. $t\in(0,T)$


The edge distributions $\hat{\ff}$ in~\EQ{aDG} must be constructed such that the consistency condition
$\hat{\ff}(\bm\mu,\bm\mu)=\ff(\bm\mu)$ holds for all $\bm\mu\in\IR^M$ and that the formulation is stable
in some appropriate sense. We propose the upwind edge distribution:
\begin{equation}
\label{eq:F+-}
\hat{\ff}(\bm\mu{}_+,\bm{\mu}_-)
=
\begin{cases}
\ff(\bm\mu{}_+)\quad&\text{if }v_{\nu}>0
\\
\ff(\bm\mu{}_-)\quad&\text{if }v_{\nu}\leq0
\end{cases}
\end{equation}
We will show that this choice of the edge distribution ensures that the DGFEM formulation~\EQ{DGform} is entropy stable in
the sense that the entropy-dissipation property~\EQ{dissipationineq} of the moment system~\EQ{ClsMomSys} with closure relation~\EQ{Cls} transfers to the DGFEM formulation.
To facilitate the derivation of the entropy-dissipation property, we
first recast~\EQ{DGform} into the equivalent form of a DGFEM/moment Galerkin approximation of the renormalized
Boltzmann equation according to~\EQ{GalLevMomCls}:
\begin{multline}
\label{eq:DGMOM}
\text{\em Find }g^h\in\mathscr{L}(0,T;V^h(\Omega,\IM)):
\\
\qquad
\int_{\Omega}
\big\langle{}w\partial_t\beta(g^h)\big\rangle
+
\sum_{e\in\mathcal{S}^h}
\int_e
\big\langle
v_{\nu}\jump{w}\hat{\beta}(g^h_+,g^h_-)\big\rangle
-
\int_{\Omega}
\big\langle{}(\partial_{x_i}w)v_i\beta(g^h)\big\rangle
=
\int_{\Omega}
\big\langle{}w\CC\big(\beta(g^h)\big)\big\rangle
\\
\qquad\forall{}w\in{}V^h(\Omega,\IM)\text{ a.e. }t\in(0,T)
\end{multline}
Our objective is to show that if $g^h$ satisfies~\EQ{DGMOM}, then~\EQ{dissipationineq} holds with $\ff$ replaced by~$\beta(g^h)$.
Note that in the boundary integral in~\EQ{dissipationineq}, $\ff$ must be replaced by~$\hat{\beta}(g^h_+,g^h_-)$.
To condense the presentation, we assume a-priori that the background distribution~$\BB$ is uniform in space and time. Similar arguments as in Section~\SEC{MomSys} then convey
\begin{equation}
\label{eq:DGd1}
\frac{d}{dt}\int_{\Omega}\big\langle\BB\,\varphi(\beta^h/\BB)\big\rangle
=
\int_{\Omega}
\big\langle{}g^h\partial_t\beta^h\big\rangle
+
\int_{\Omega}
\int_{\mathrm{supp}^c(\beta^h)}
(\varphi'(0)-g^h)\partial_t\beta^h
\end{equation}
with the abridged notation $\beta^h:=\beta(g^h)$. Noting that $g^h(t)$ resides in $V^h(\Omega,\IM)$ a.e. $t\in(0,T)$,
we can apply Galerkin orthogonality according to~\EQ{DGMOM} to recast the first term in~\EQ{DGd1} into:
\begin{equation}
\label{eq:DGd2}
\int_{\Omega}
\big\langle{}g^h\partial_t\beta^h\big\rangle
=
-
\sum_{e\in\mathcal{S}^h}
\int_e
\big\langle
v_{\nu}\jump{g^h}\hat{\beta}{}^h\big\rangle
+
\sum_{\kappa\in\mathcal{T}^h}
\int_{\kappa}
\big\langle{}(\partial_{x_i}g^h)v_i\beta^h\big\rangle
+
\sum_{\kappa\in\mathcal{T}^h}
\int_{\kappa}
\big\langle{}g^h\CC(\beta^h)\big\rangle
\end{equation}
with $\hat{\beta}{}^h:=\hat{\beta}(g^h_+,g^h_-)$. The second term in~\EQ{DGd2} can be recast into
\begin{equation}
\label{eq:DGd2a}
\begin{aligned}
\sum_{\kappa\in\mathcal{T}^h}
\int_{\kappa}
\big\langle{}(\partial_{x_i}g^h)v_i\beta^h\big\rangle
&=
\sum_{e\in\mathcal{S}^h}\big\langle{}v_{\nu}\jump{g^h\beta^h}\big\rangle
-
\sum_{\kappa\in\mathcal{T}^h}
\int_{\kappa}
\big\langle{}g^hv_i\partial_{x_i}\beta^h\big\rangle
\end{aligned}
\end{equation}
The second term in the right member of~\EQ{DGd2a} can in turn be reformulated as
\begin{equation}
\label{eq:DGd3a}
\begin{aligned}
\sum_{\kappa\in\mathcal{T}^h}
\int_{\kappa}\big\langle{}g^hv_i\partial_{x_i}\beta^h\big\rangle
&=
\sum_{\kappa\in\mathcal{T}^h}
\int_{\kappa}
\big\langle{}\varphi'(\beta^h/\BB)v_i\partial_{x_i}\beta^h\big\rangle
-
\sum_{\kappa\in\mathcal{T}^h}
\int_{\kappa}
\int_{\mathrm{supp}^c(\beta^h)}(\varphi'(0)-g^h)v_i\partial_{x_i}\beta^h
\\
&=
\sum_{\kappa\in\mathcal{T}^h}
\int_{\kappa}\partial_{x_i}\big\langle{}v_i\BB\,\varphi(\beta^h/\BB)\big\rangle
-
\sum_{\kappa\in\mathcal{T}^h}
\int_{\kappa}
\int_{\mathrm{supp}^c(\beta^h)}(\varphi'(0)-g^h)v_i\partial_{x_i}\beta^h
\\
&=
\sum_{e\in\mathcal{S}^h}
\big\langle{}v_{\nu}\jump{\BB\,\varphi(\beta^h/\BB)}\big\rangle
-
\sum_{\kappa\in\mathcal{T}^h}
\int_{\kappa}
\int_{\mathrm{supp}^c(\beta^h)}(\varphi'(0)-g^h)v_i\partial_{x_i}\beta^h
\end{aligned}
\end{equation}
%
Collecting the results in Equations~\EQ{DGd1}-\EQ{DGd3a} and recalling that
$g=\varphi'(\beta(g)/\BB)$ on the support of~$\beta$, we obtain
\begin{equation}
\label{eq:DGd5}
\frac{d}{dt}\int_{\Omega}\big\langle\BB\,\varphi(\beta^h/\BB)\big\rangle
+\int_{\partial\Omega}\big\langle{}v_n\BB\,\varphi(\hat{\beta}^h/\BB)\big\rangle=T_{\mathcal{S}^h}+T_{\mathcal{T}^h}
\end{equation}
with
\begin{align}
T_{\mathcal{S}^h}
&=
\sum_{e\in\mathcal{S}^h}
\Big\langle v_{\nu}
\bjump{\varphi'(\beta^h/\BB)(\beta^h-\hat{\beta}{}^h)-\big(\BB\varphi(\beta^h/\BB)-\BB\varphi(\hat{\beta}^h/\BB)\big)}
\Big\rangle
\notag
\\
&\qquad\qquad
-
\sum_{e\in\mathcal{S}^h}
\bigg[\!\!\bigg[
\int_{\mathrm{supp}^c(\beta^h)}v_{\nu}(\varphi'(0)-g^h)(\beta^h-\hat{\beta}{}^h)
\bigg]\!\!\bigg]
\label{eq:TSh}
\\
T_{\mathcal{T}^h}
&=
\sum_{\kappa\in\mathcal{T}^h}
\int_{\kappa}
\big\langle{}\varphi'(\beta^h/\BB)\CC(\beta^h)\big\rangle
+
\sum_{\kappa\in\mathcal{T}^h}
\int_{\kappa}
\int_{\mathrm{supp}^c(\beta^h)}(\varphi'(0)-g^h)(\partial_t\beta^h+v_i\partial_{x_i}\beta^h-\CC(\beta^h))
\end{align}
Let us note that $\jump{\BB\varphi(\hat{\beta}^h/\BB)}$ in~\EQ{TSh} vanishes on interior edges by virtue of
the continuity of $\hat{\beta}^h$, and that its aggregated contribution coincides with the boundary integral in the
left member of~\EQ{DGd5}. Similar arguments as in Section~\SEC{MomSys} convey that $\smash[tb]{T_{\mathcal{T}^h}\leq{}0}$; cf.~\EQ{T1}.
To assess the contribution of~$\smash[tb]{T_{\mathcal{S}^h}}$, we recall that $\smash[tb]{\hat{\beta}^h=\beta^h_+}$
(resp.~$\smash[tb]{\hat{\beta}^h=\beta^h_-}$)
if $v_{\nu}>0$ (resp. $v_{\nu}\leq{}0$). For $v_{\nu}>0$ it therefore follows that
\begin{equation}
\label{eq:jump1}
\bjump{\varphi'(\beta^h/\BB)(\beta^h-\hat{\beta}{}^h)-\big(\BB\varphi(\beta^h/\BB)-\BB\varphi(\hat{\beta}^h/\BB)\big)}
=
-
\big(\varphi'(\beta^h_-/\BB)(\beta^h_--\hat{\beta}{}^h)-\big(\BB\varphi(\beta^h_-/\BB)-\BB\varphi(\hat{\beta}^h/\BB)\big)\big)
\end{equation}
From the convexity of $\BB\varphi((\cdot)/\BB)$ we infer that for any $\beta,\hat{\beta}\in\IR$:
\begin{equation}
\varphi'(\beta/\BB)(\beta-\hat{\beta})
\geq
\BB\varphi(\beta/\BB)-\BB\varphi(\hat{\beta}/\BB)
\end{equation}
Hence, for $v_{\nu}>0$ the jump term in the left member of~\EQ{jump1} is non-positive. Similarly, one can infer that
the jump term is non-negative if $v_{\nu}\leq{}0$. The first term in~$T_{\mathcal{S}^h}$ is therefore non-positive.
The contribution of each edge to the second term in~$T_{\mathcal{S}^h}$ can be decomposed as:
\begin{multline}
\label{eq:DGdfin}
\bigg[\!\!\bigg[
\int_{\mathrm{supp}^c(\beta^h)}v_{\nu}(\varphi'(0)-g^h)(\beta^h-\hat{\beta}{}^h)
\bigg]\!\!\bigg]
=
\int_{\mathrm{supp}^c(\beta^h_+),v_{\nu}\leq{}0}v_{\nu}(\varphi'(0)-g^h_+)(\beta^h_+-\beta^h_-)
\\
-
\int_{\mathrm{supp}^c(\beta^h_-),v_{\nu}>0}v_{\nu}(\varphi'(0)-g^h_-)(\beta^h_--\beta^h_+)
\end{multline}
It generally holds that $\varphi'(0)\geq{}g^h_{\pm}$; cf.~\EQ{psi}. In the first term in~\EQ{DGdfin}
the domain of integration is a subset of $\mathrm{supp}^c(\beta^h_+)$ and accordingly $\beta^h_+$ vanishes.
It then follows from $v_{\nu}\leq{}0$ and $\beta^h_-\geq{}0$ that this term is non-negative. In a similar manner,
it can be shown that the contribution of the second term in~\EQ{DGdfin} is also non-negative. Both terms in~\EQ{TSh}
are therefore non-positive.

\section{Numerical results}
\label{sec:NumRes}
To illustrate the properties of the proposed discontinuous Galerkin finite-element moment method~\EQ{DGMOM}, or equivalently~\EQ{DGform}, we present numerical experiments for a one-dimensional shock-tube problem. Before doing so, the problem specification and moment-system approximation must be completed by specifying the collision operator and closure relation. We restrict ourselves here to the standard BGK collision operator~\cite{Bhatnagar:1954hc}, viz.
$\CC(f) = -\tau^{-1}(f-\MM_f)$ with $\MM_f$ the local Maxwellian with the same invariant moments
as~$f$ and~$\tau^{-1}$ a relaxation rate.
We adopt the relaxation parameter in accordance with the hard-sphere collision process of Bird~\cite{Bird:1994}: $\tau = (5\lambda/16)(2\pi\rho/p)^{1/2}$ with $\lambda$ the mean free path.
We identify the Knudsen number with the mean free path divested of its units. The closure relation is defined by the renormalization map:
\begin{equation}
\label{eq:jspCls}
\beta(g) = \BB\left(1+\frac{g}{N}\right)^N_+,
\end{equation}
where $(\cdot)_+ = \frac{1}{2}(\cdot)+\frac{1}{2}|\cdot|$ is the non-negative part of a function extended by~$0$ and $N\in\mathbb{N}$; see~\cite{Abdel-Malik:2015eu} for further details on this closure relation. The corresponding $\varphi$-divergence based entropy is $f\mapsto\langle{}\BB\varphi(f/\BB)\rangle$ with~$\varphi$ according to~\EQ{varphiN}, i.e.~\EQ{jspCls} corresponds to the solution of the constrained entropy-minimization problem~\EQ{EntMin} with $\varphi$ from~\EQ{varphiN}. In~\cite{Abdel-Malik:2015eu} it is shown that the minimization problem in~\EQ{EntMin} is well-posed for this entropy. Moreover, $\langle\BB\varphi(f/\BB)\rangle$ corresponds to an entropy for the BGK operator in compliance with the dissipation relation~\EQ{Dissipation};
see also~\cite{Abdel-Malik:2015eu}. In the sequel, we set~$N=2$.
%

We regard Sod's shock-tube problem on a spatial domain $x\in(-1,1)$ and a time interval $t\in(0,0.1)$. This test case concerns a Riemann problem with discontinuous initial data corresponding to a piecewise uniform Maxwellian distribution:
\begin{equation}
\label{eq:ICdata}
  f(0,x,v)=
  \begin{cases}
    f_L(v) = \mathcal{M}_{(1,0,1)}(v) \qquad &\text{for } x\leq0 \\
    f_R(v) = \mathcal{M}_{(\frac{1}{2},0,\frac{2}{5})}(v) \qquad &\text{for } x>0
  \end{cases}
\end{equation}
cf.~\EQ{Maxwellian}; see~\cite{McDonald:2013uq}. Two different approximations are considered. For the first approximation we regard a uniform background distribution $\BB = \mathcal{M}_{(1,0,1)}$. In the second approximation the background distribution corresponds to the solution of the Vlasov equation with initial data~\EQ{ICdata}, locally regularized near~$x=0$:
\begin{subequations}
\label{eq:FreeStreamB}
  \begin{align}
    \partial_t \BB + \partial_x v\BB &= 0 \\
    \BB(0,x,v) = \BB_0(x,v) &=
    \begin{cases}
      \mathcal{M}_{(1,0,1)}(v) \qquad &\text{for } x\leq x_0\\
      \frac{x_1-x}{x_1-x_0}\mathcal{M}_{(1,0,1)}(v)  +  \frac{x-x_0}{x_1-x_0}\mathcal{M}_{(\frac{1}{2},0,\frac{2}{5})}(v)  \qquad &\text{for } x_0<x<x_1\\
      \mathcal{M}_{(\frac{1}{2},0,\frac{2}{5})}(v) \qquad &\text{for } x\geq x_1
    \end{cases}
  \end{align}
\end{subequations}
with $x_1=-x_0=\frac{1}{64}$. The regularization near $x=0$ serves to avoid
complications related to discontinuities in the background distribution; see Section~\SEC{MomSys}. The solution to (\ref{eq:FreeStreamB}) is given by $\BB(t,x,v)=\BB_0 (x-vt,v)$.
The parameters $\rho,p$ in the relaxation parameter~$\tau$ are determined
from~$f_L$ in the initial distribution~\EQ{ICdata}.

We restrict ourselves here to finite-element approximation spaces of polynomial
degree~$p=0$, i.e. element-wise constant approximations in position dependence.
For the time-integration procedure, we apply the forward Euler scheme with a time
step that is sufficiently small to render the numerical results essentially independent of the time step.

\begin{figure}
\begin{center}
\includegraphics[width=0.45\textwidth]{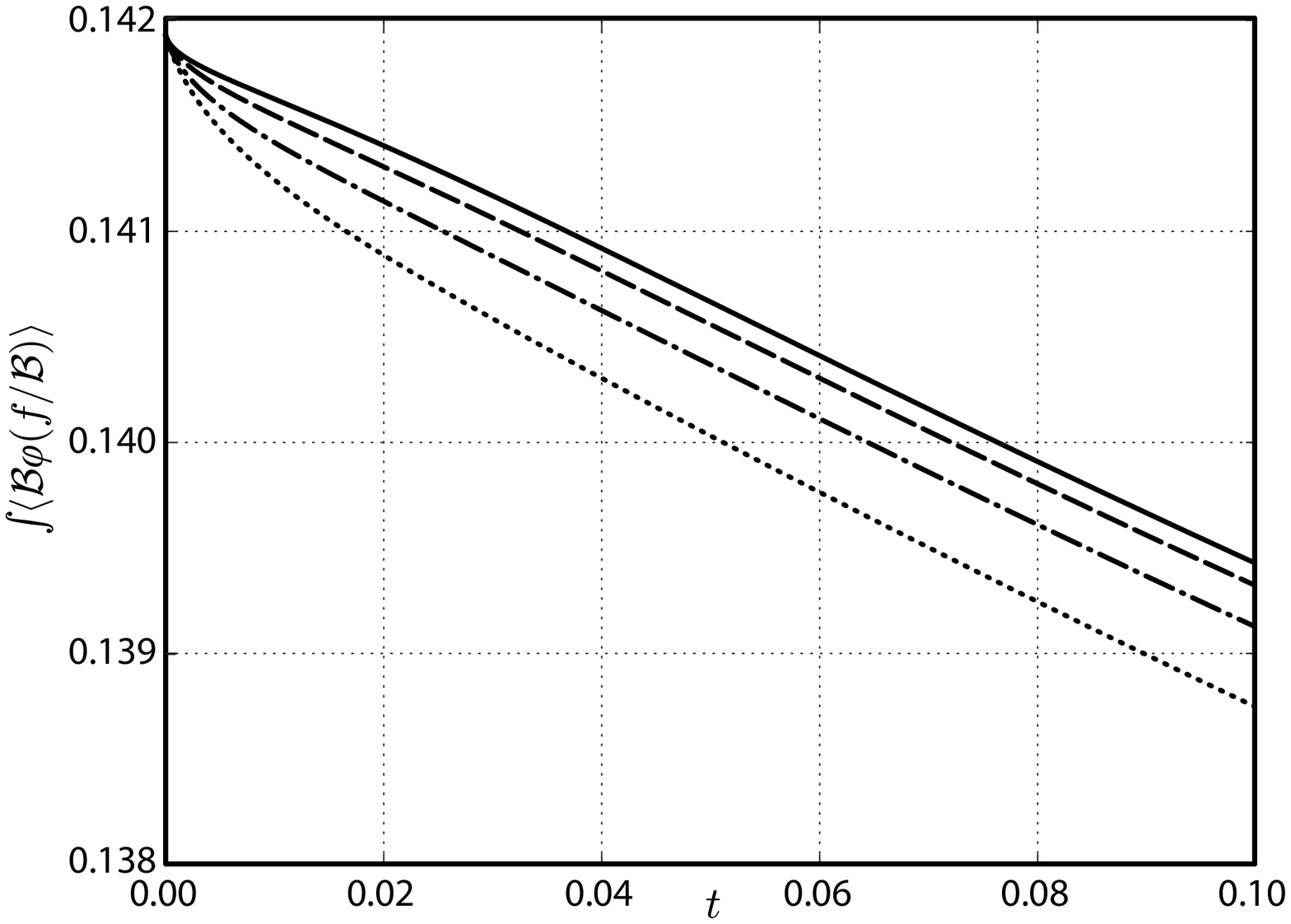}
\hspace{10pt}
\includegraphics[width=0.45\textwidth]{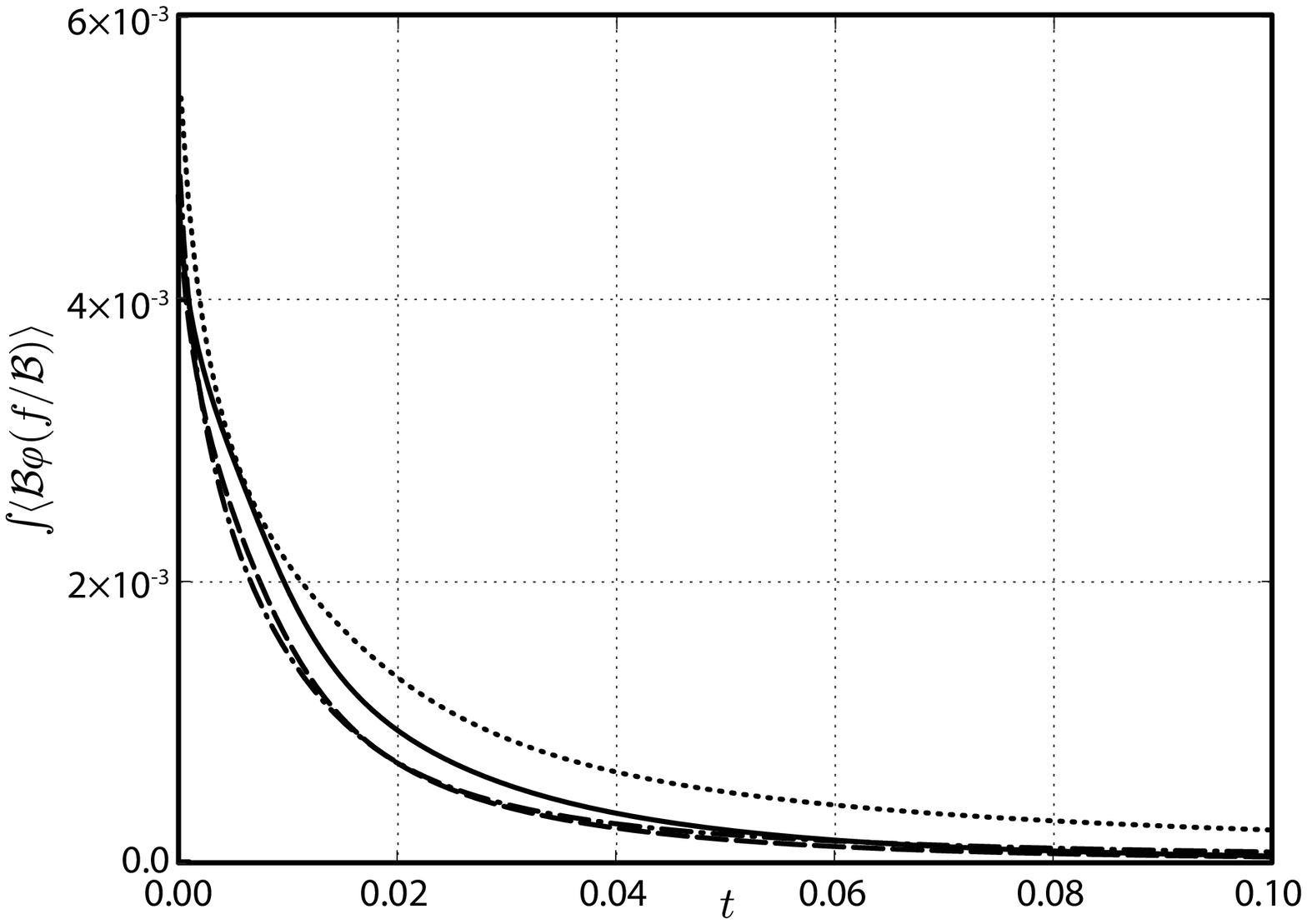}
\\
\includegraphics[width=0.45\textwidth]{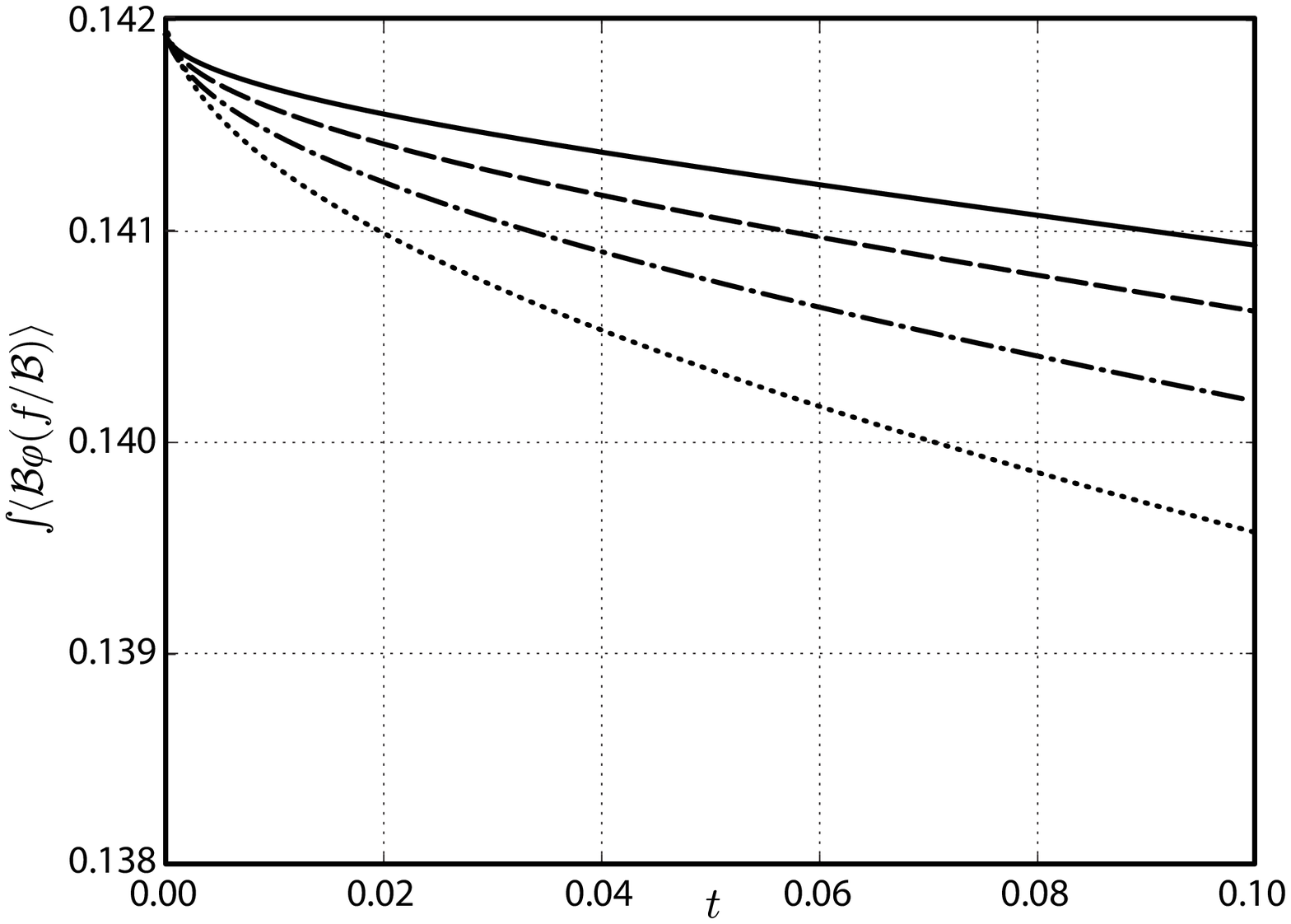}
\hspace{5pt}
\includegraphics[width=0.45\textwidth]{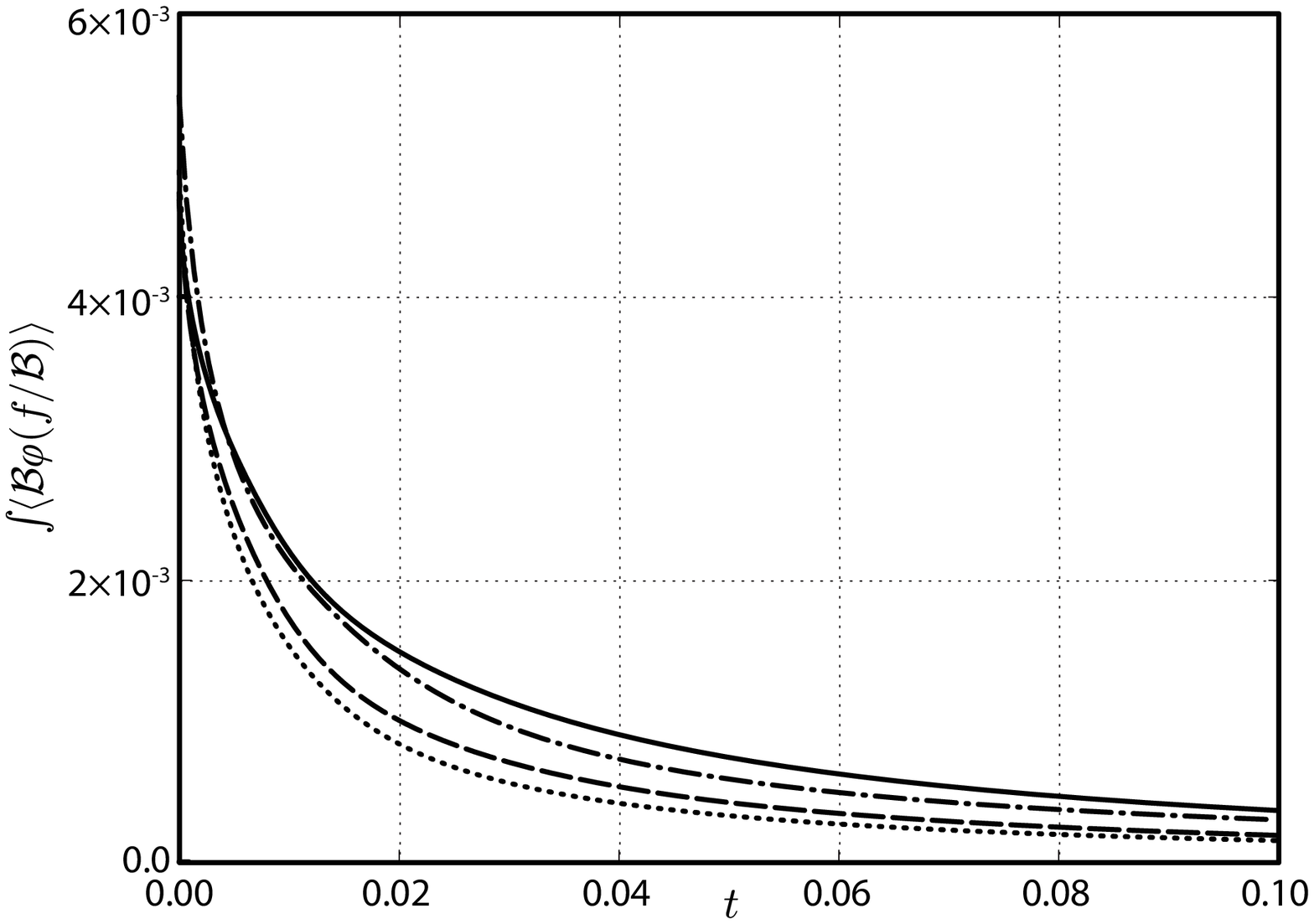}
\\
\includegraphics[width=0.45\textwidth]{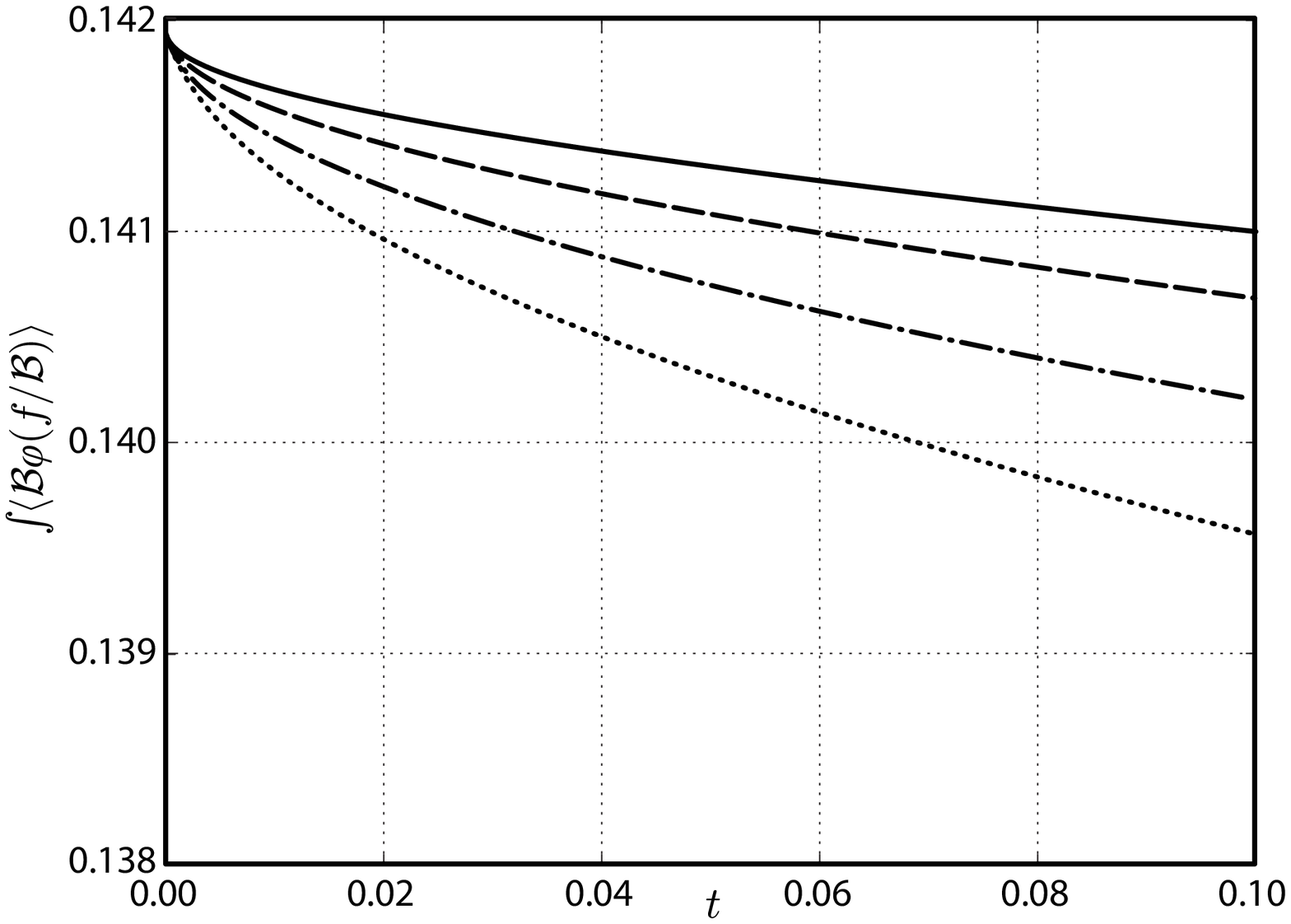}
\hspace{10pt}
\includegraphics[width=0.45\textwidth]{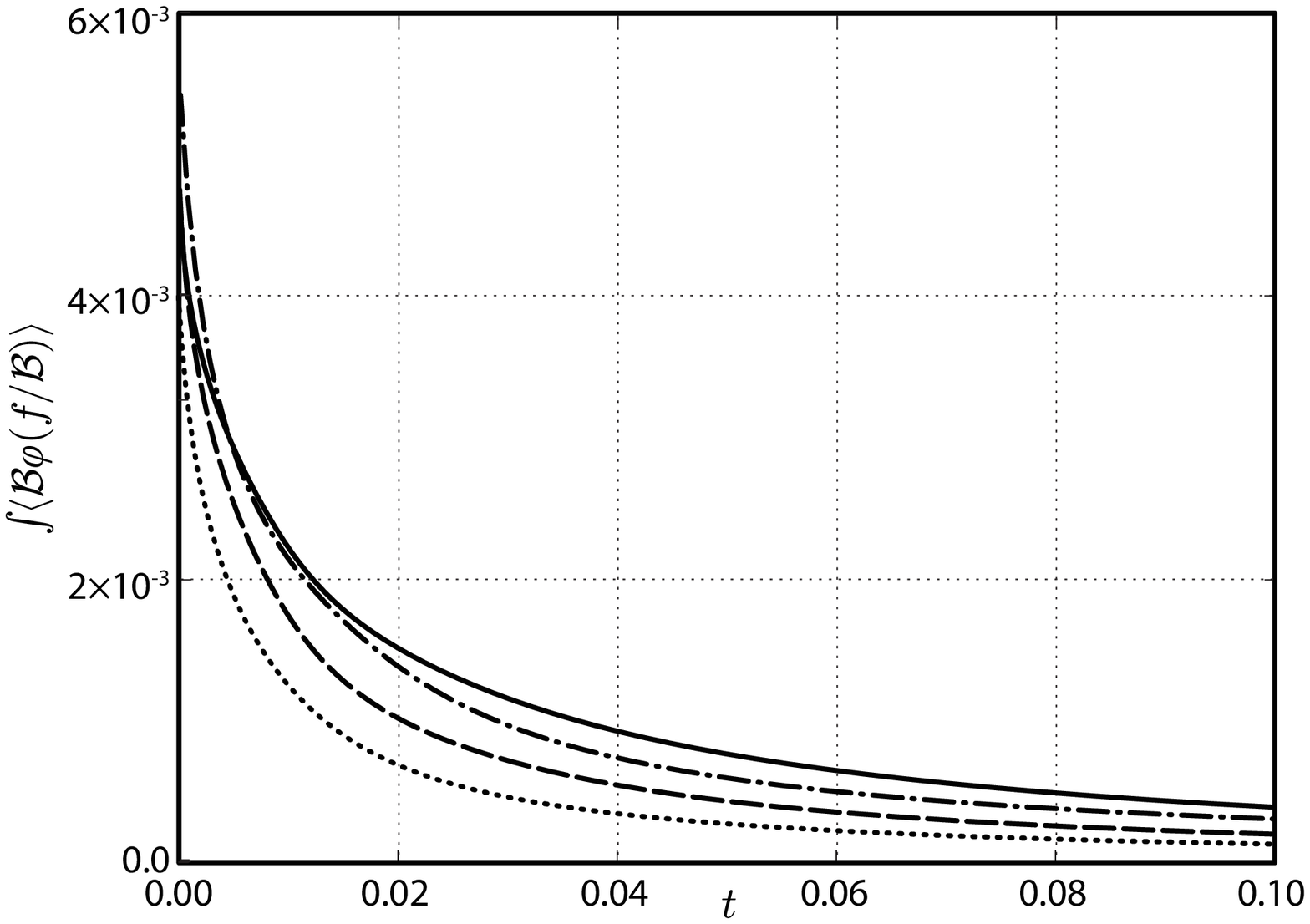}
\end{center}
\caption{Evolution of the entropy $\int_{\Omega}\big\langle\BB\,\varphi(\ff/\BB)\big\rangle$ for
the DGFEM 5-moment approximation, for the uniform background distribution $\BB=\IM_{(1,0,1)}$
({\em left\/}) and the non-uniform background distribution in~\EQ{FreeStreamB}
({\em right\/}), for Knudsen numbers
$\Kn=2.3\times{}10^{-2}$ ({\em top\/}),
$\Kn=2.3\times{}10^{0}$ ({\em center\/})
and
$\Kn=2.3\times{}10^{+1}$ ({\em bottom\/}), and for uniform meshes with mesh width
$h=2^{-7}$ (\,\protect\rule[1.5pt]{1pt}{1pt} \protect\rule[1.5pt]{1pt}{1pt}\,),
$h=2^{-8}$ (\,\protect\rule[1.5pt]{8pt}{1pt} \protect\rule[1.5pt]{1pt}{1pt}\,),
$h=2^{-9}$ (\,\protect\rule[1.5pt]{8pt}{1pt} \protect\rule[1.5pt]{8pt}{1pt}\,)
and $h=2^{-10}$ (\,\protect\rule[1.5pt]{12pt}{1pt}\,).
\label{fig:Entropy}}
\end{figure}

Figure~\FIG{Entropy} displays the evolution of the entropy $\int_{\Omega}\big\langle\BB\,\varphi(\ff/\BB)\big\rangle$ for the 5-moment system, i.e. for $\IM$ spanned
by $\{1,v,\ldots,v^4\}$, for both the uniform and non-uniform
background distributions, for Knudsen numbers
$\Kn\in{}2.3\times{}10^{\{-2,0,+1\}}$, and for uniform meshes with mesh width
$h\in{}2^{\{-7,\ldots,-10\}}$. The results in Figure~\FIG{Entropy} corroborate the
entropy-dissipation property of the DGFEM moment method.
Observing that the results for Knudsen numbers $\Kn\in2.3\times{}10^{\{0,+1\}}$ are nearly identical,
we infer that in the corresponding highly rarefied and transition regimes the entropy dissipation
is dominated by the dissipation induced by the discontinuities in the DG approximation,
for the considered sequence of finite-element spaces. This is also evident from the
deviation between the graphs for distinct~$h$. In the near continuum
regime $\Kn=2.3\times{}10^{-2}$ and on sufficiently fine meshes, the effect of
the discontinuities on the entropy dissipation is less pronounced.

\begin{figure}
\begin{center}
\includegraphics[width=0.45\textwidth]{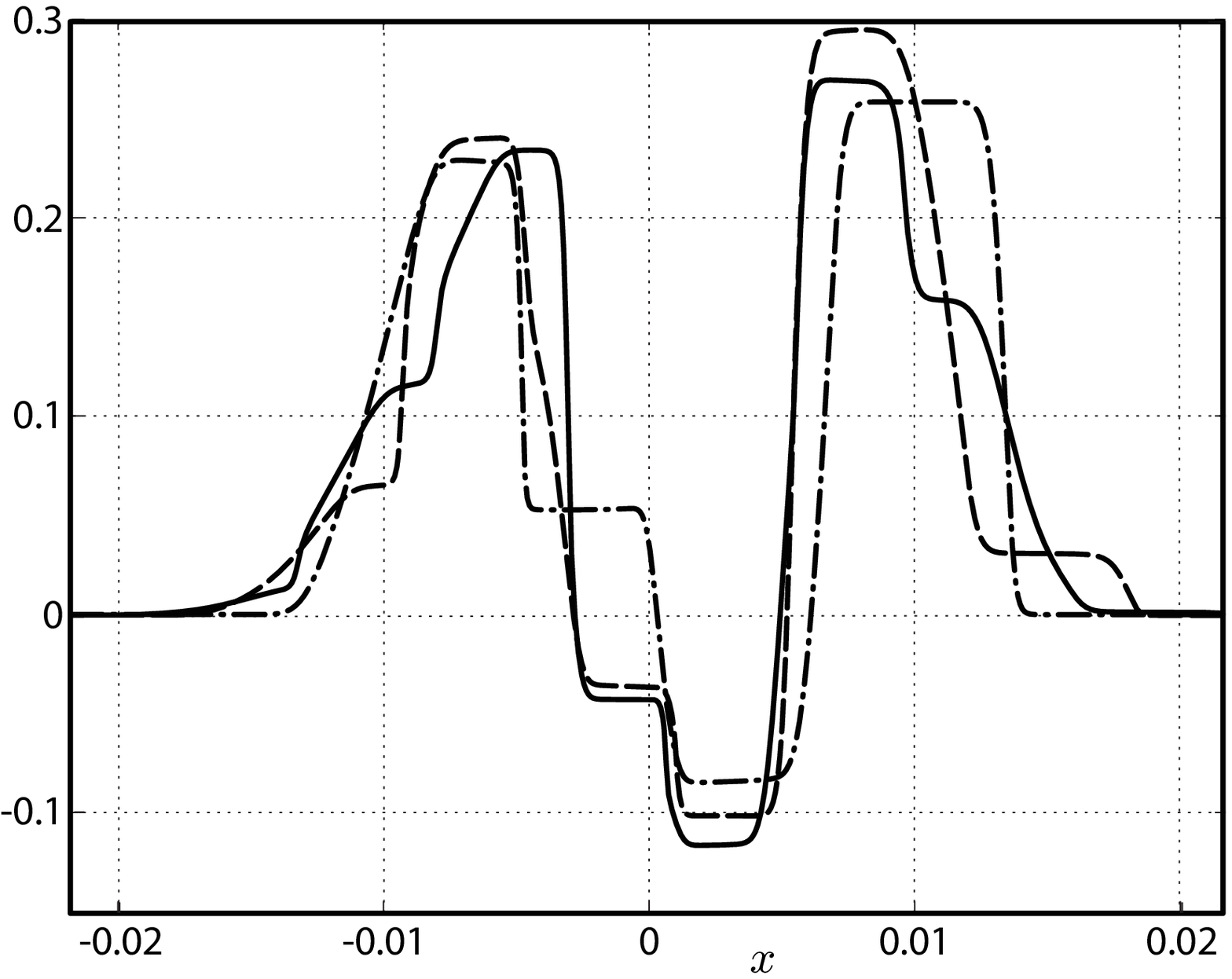}
\hspace{10pt}
\includegraphics[width=0.45\textwidth]{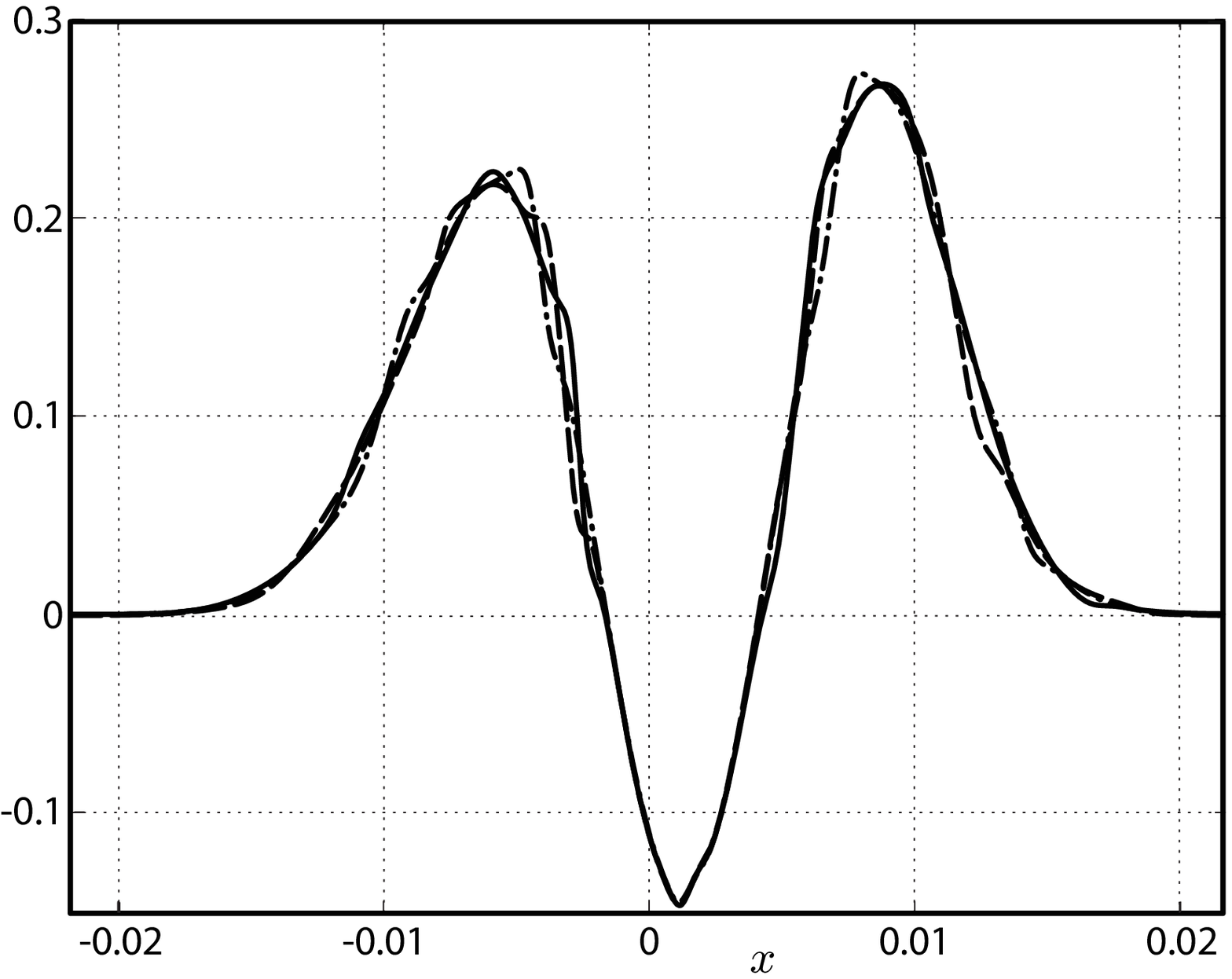}
\end{center}
\caption{
Heat flux $q^{\star}$ according to~\EQ{qstar} for $\Kn=2.3\times{}10^{+1}$ at $t=1/10$,
for the uniform background distribution $\BB=\IM_{(1,0,1)}$
({\em left\/}) and the non-uniform background distribution in~\EQ{FreeStreamB}
({\em right\/}), for
$M=5$ (\,\protect\rule[1.5pt]{8pt}{1pt} \protect\rule[1.5pt]{1pt}{1pt}\,),
$M=7$ (\,\protect\rule[1.5pt]{8pt}{1pt} \protect\rule[1.5pt]{8pt}{1pt}\,)
and
$M=9$ (\,\protect\rule[1.5pt]{12pt}{1pt}\,)
moments and mesh width $h=2^{-10}$.
\label{fig:Heat}}
\end{figure}

Figure~\FIG{Heat} presents the normalized heat flux according to
\begin{equation}
\label{eq:qstar}
q^{\star}=\rho^{1/2}p^{-3/2}q
\end{equation}
 with $\rho=\langle{}f\rangle$ as density, $p=\tfrac{1}{2}\langle(u-v)^2f\rangle$ as pressure,
 $q=\langle{}(v-u)^3f\rangle$ as heat flux and $u=\langle{}vf\rangle/\langle{}f\rangle$ as the
 macroscopic velocity,
in the highly-rarefied regime ($\Kn=2.3\times{}10^{+1}$) at $t=1/10$, extracted from the DGFEM moment-method approximation for $M\in\{5,7,9\}$ and mesh width $h=2^{-10}$, for both the uniform and non-uniform background distribution. Let us note that
heat transfer is a typical non-equilibrium effect that is not represented by the conventional
continuum model of gas dynamics, viz. the Euler equations with ideal-gas closure. The results
in Figure~\FIG{Heat} convey that the approximation of the heat flux for the uniform background distribution
is still relatively sensitive to the number of moments that is applied in the approximation. By virtue of the high Knudsen number, the background distribution~\EQ{FreeStreamB} derived from the Vlasov solution yields an accurate approximation to the actual solution of the Boltzmann
equation, which translates into very high accuracy of the corresponding heat-flux approximation,
virtually independent of the number of moments. It is to be noted, however, that the accuracy of the heat-flux approximation pertaining to the uniform background distribution in
Figure~\FIG{Heat}~({\em left\/}) also compares favorably to corresponding results in the literature for other closure relations; cf. for instance~\cite{McDonald:2013uq}.

\section{Conclusion}
\label{sec:Conc}
We have presented a Galerkin approximation method for the Boltzmann equation based
on the combination of moment-system approximation in velocity dependence and discontinuous-Galerkin
finite-element approximation in position dependence. For the moment systems, we considered
a closure relation that derives from the minimization of a relative entropy corresponding
to a $\varphi$\nobreakdash-divergence. The background measure in the relative entropy then appears
as a factor in the approximation of the one-particle marginal. We established that for suitable
background measures, the moment systems retain the fundamental structural properties of the
underlying Boltzmann equation, viz., conservation of mass, momentum and energy, Galilean invariance and dissipation of a relative entropy. Moreover, the divergence-based closure leads to a hierarchy of tractable symmetric hyperbolic systems. The moment systems can alternatively be conceived of as Galerkin approximations of the Boltzmann equation in renormalized form.

For the discretization of the moment systems in position dependence, we considered the discontinuous Galerkin finite-element method. The combined DGFEM moment method can be construed as a Galerkin finite-element approximation of the Boltzmann equation in renormalized form, based on a
tensor-product approximation space composed of the DGFEM approximation space in position dependence and global polynomials in velocity dependence. We introduced a new class of numerical flux functions for the combined DGFEM moment method. This new numerical flux function appears naturally in the setting of the renormalized Boltzmann equation as the upwind distribution at the inter-element interfaces. We established that this upwind flux renders the DGFEM moment method entropy stable, i.e. the entropy-dissipation property of the
moment systems transfers to the DGFEM formulation.

Numerical results were presented for a one-dimensional shock-tube problem in the highly-rarefied, intermediate and near-continuum regimes. We considered two different approximations: one based on a background distribution corresponding to a uniform Maxwellian, and one based on a background distribution corresponding to a solution to the Vlasov equation. The numerical results confirm the entropy-dissipation property of the DGFEM moment method for both types of approximation. In the highly rarefied regime, the approximation based on the Vlasov background distribution provides
excellent approximations of the heat flux, essentially independent of the number of moments. The approximation corresponding to the uniform background distribution yields very accurate results for the heat flux, that compare favorably to corresponding results in the literature for other closure
 relations.

\bibliography{BibFile}   
\end{document}